# Derivative of Rotation Matrix – Direct Matrix Derivation of Well-Known Formula

Fumio Hamano
Department of Electrical Engineering
California State University, Long Beach
Long Beach, CA USA
Fumio.Hamano@csulb.edu

*Abstract*—In motion Kinematics, it is well-known that the time derivative of a $3 \times 3$ rotation matrix equals a skew-symmetric matrix multiplied by the rotation matrix where the skew symmetric matrix is a linear (matrix-valued) function of the angular velocity and the rotation matrix represents the rotating motion of a frame with respect to a reference frame. The equation is widely used in engineering, e.g., robotics, control, air/spacecraft modeling, etc. However, the derivations found in the literature are indirect. Motivated by the fact that the set of $3 \times 3$ rotation matrices, i.e., SO(3), is a Lie group, forming a smooth (differentiable) manifold, we describe the infinitesimal increment of the rotation matrix in terms of rotation matrices and show that the above equation immediately follows.

## I. INTRODUCTION

In motion Kinematics, it is well-known that the time derivative of the $3 \times 3$ rotation matrix equals a skew-symmetric matrix multiplied by the rotation matrix where the skew symmetric matrix is a linear (matrix-valued) function of the angular velocity and the rotation matrix represents the rotating motion of a frame with respect to a reference frame. The equation is widely used in engineering, e.g., robotics, control, air/spacecraft modeling, etc. (See [1], [2], [3], etc.) However, the derivations found in the literature are indirect. (See [1] Ch. 4 and [2] Ch. 3, for instance.) Motivated by the fact that the set of $3 \times 3$ rotation matrices, i.e., SO(3, $\mathbb{R}$), is a Lie group, forming a smooth (differentiable) manifold (see [4]), we describe the infinitesimal increment of the rotation matrix in terms of rotation matrices using composition rule as reference to fixed reference frame and show that the above equation immediately follows. The derivation is constructive and explicit.

In Sec. II the rotation matrix is defined and a composition rule as reference to a "fixed" frame is reviewed. In Sec. III infinitesimal rotation and angular velocity are described. And an alternative direct derivation of the derivative equation for a rotation matrix is given in Sec. IV, which is the main section of this paper. The concluding remarks are provided in Sec. V.

## II. ROTATION MATRIX AND ITS COMPOSITION

### A. Definition of Rotation Matrix

Let us consider 3-dimensional Cartesian space with two Cartesian coordinate frames: frame 0 (or reference frame) and frame 1. (See Fig. 1.) The frame 1 rotates relative to frame 0. We define the $3 \times 3$ rotation matrix relating frame 1 to frame 0 as follows:

$$R_1^0 := \begin{bmatrix} \mathbf{i}_1 \cdot \mathbf{i}_o & \mathbf{j}_1 \cdot \mathbf{i}_o & \mathbf{k}_1 \cdot \mathbf{i}_o \\ \mathbf{i}_1 \cdot \mathbf{j}_o & \mathbf{j}_1 \cdot \mathbf{j}_o & \mathbf{k}_1 \cdot \mathbf{j}_o \\ \mathbf{i}_1 \cdot \mathbf{k}_o & \mathbf{j}_1 \cdot \mathbf{k}_o & \mathbf{k}_1 \cdot \mathbf{k}_o \end{bmatrix} \in SO(3, \mathbb{R}) \qquad (1)$$

where $\mathbf{i}_o, \mathbf{j}_o, \mathbf{k}_o, \mathbf{i}_1, \mathbf{j}_1,$ and $\mathbf{k}_1$ are respectively the unit vectors along the $x_o, y_o, z_o, x_1, y_1,$ and $z_1$ axes (see Fig. 1) and each element of the matrix is written as a dot product of two vectors.

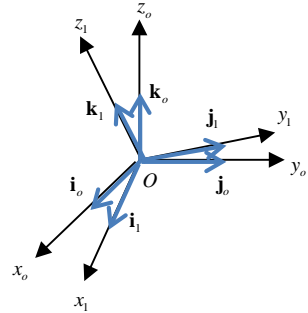

Fig. 1. Cartesian coordinate frames 0 and 1.

Since each column of $R_1^0$ represents $\mathbf{i}_1$, $\mathbf{j}_1$, and $\mathbf{k}_1$ in frame 0, we may interpret the matrix as a representation of frame 1 as reference to frame 0. We will use $R$ in place of coordinate frame specific notation such as $R_1^0$ since the rotation matrix is the same as long as the relative orientation of two frames is the same. (Extensive discussions related to rotation matrices can be found in [1] Ch.2], [2] Ch. 2, and [5] Ch.2.

## B. Composition of Rotations with Respect to (Fixed) Reference Frame

If the rotation matrix $R$ represents the orientation of frame 1 as a result of two consecutive rotations with respect to the fixed reference frame (frame 0), it is well-known (see [1] Ch. 2; [2] Ch. 2) that $R$ is described by the product of two rotation matrices representing the individual rotations in the reverse order, i.e.,

$$R = R_{2nd} R_{1st} \qquad (2)$$

where $R_{1st}$ and $R_{2nd}$ are respectively the rotation matrices representing the first and $2^{nd}$ rotations.

## III. INFINITESIMAL ROTATION AND ANGULAR VELOCITY

If the new frame (say, frame 1) is obtained from the reference frame (say, frame 0) by consecutive rotations about the $x$-, $y$-, and $z$-axes respectively by infinitesimal amounts $d\phi_x$, $d\phi_y$, and $d\phi_z$ in any order, it is easy to see by calculation that the combined rotation $R(d\phi_x, d\phi_y, d\phi_z)$ is represented by

$$R(d\phi_x, d\phi_y, d\phi_z) = \begin{bmatrix} 1 & -d\phi_z & d\phi_y \\ d\phi_z & 1 & -d\phi_x \\ -d\phi_y & d\phi_x & 1 \end{bmatrix} \qquad (3)$$

Since (3) does not depend on the order of the individual three rotations, the triple $(d\phi_x, d\phi_y, d\phi_z)$ defines the orientation of frame 1 as reference to frame 0. We define

$$d\phi := \begin{bmatrix} d\phi_x \\ d\phi_y \\ d\phi_z \end{bmatrix} \qquad (4)$$

and call it infinitesimal rotation. Trivially, $d\phi$ is scalable, i.e., a constant multiple of the triple in (4) results in another triple of infinitesimal angles, which represent another infinitesimal rotation of the frame. The quantity $d\phi$ also admits addition, as shown below. Thus, $d\phi$ can be treated as a vector with usual scalar multiplication and addition of column vectors (all vector space axioms are satisfied). The validity of the addition can be seen as follows. Let $d\tilde{\phi}_x$, $d\tilde{\phi}_y$, and $d\tilde{\phi}_z$ be another set of infinitesimal angles of rotations about the $x$-, $y$-, and $z$-axes, respectively. It is easy to verify by calculation the following composition rule:

$$R(d\phi_x, d\phi_y, d\phi_z) R(d\tilde{\phi}_x, d\tilde{\phi}_y, d\tilde{\phi}_z)$$
$$= R(d\tilde{\phi}_x, d\tilde{\phi}_y, d\tilde{\phi}_z) R(d\phi_x, d\phi_y, d\phi_z) \qquad (5)$$
$$= R(d\phi_x + d\tilde{\phi}_x, d\phi_y + d\tilde{\phi}_y, d\phi_z + d\tilde{\phi}_z)$$

The angular velocity $\omega$ is defined by

$$\omega := \frac{d\phi}{dt} = \begin{bmatrix} d\phi_x / dt \\ d\phi_y / dt \\ d\phi_z / dt \end{bmatrix} \qquad (6)$$

The detailed discussion on infinitesimal rotation and angular velocity can be found in [5] Sec.'s 3.1.2 and 3.1.3.

## IV. INFINITESIMAL CHANGE OF ROTATION MATRIX DUE TO INFINITESIMAL ROTATION AND DERIVATIVE OF ROTATION MATRIX

Let $d\phi$ be the infinitesimal rotation relative to the (fixed) reference frame as defined by (4). We denote by $dr_{ij}$ the corresponding increment of the $(i,j)$-element of $R$ where $i, j = 1, 2, 3$. We define the infinitesimal change $dR$ of rotation matrix $R$ due to $d\phi$ by the $3 \times 3$ matrix with elements $dr_{ij}$'s. Then, we have

$$dR = R(d\phi)R - R$$
$$= (R(d\phi) - I)R \qquad (7)$$
$$= S(d\phi)R$$

where $S(d\phi)$ is a skew-symmetric matrix defined by

$$S(d\phi) := \begin{bmatrix} 0 & -d\phi_z & d\phi_y \\ d\phi_z & 0 & -d\phi_x \\ -d\phi_y & d\phi_x & 0 \end{bmatrix} \qquad (8)$$

Dividing $dR$ by infinitesimal time increment $dt$, we obtain

$$\frac{dR}{dt} = \frac{1}{dt} S(d\phi) R$$
$$= \begin{bmatrix} 0 & -d\phi_z/dt & d\phi_y/dt \\ d\phi_z/dt & 0 & -d\phi_x/dt \\ -d\phi_y/dt & d\phi_x/dt & 0 \end{bmatrix} R \qquad (9)$$
$$= \begin{bmatrix} 0 & -\omega_z & \omega_y \\ \omega_z & 0 & -\omega_x \\ -\omega_y & \omega_x & 0 \end{bmatrix} R$$
$$= S(\omega) R$$

where

$$S(\omega) := \begin{bmatrix} 0 & -\omega_z & \omega_y \\ \omega_z & 0 & -\omega_x \\ -\omega_y & \omega_x & 0 \end{bmatrix} \qquad (10)$$

Thus, we have derived the well-known equation

$$\frac{dR}{dt} = S(\omega) R \qquad (11)$$

where $S(\omega)$ is the skew-symmetric matrix defined by (10).

## V. CONCLUSION

We have derived the well-known rotational motion kinematic equation (11) by describing the infinitesimal increment $dR$ in terms of rotation matrices and by using a composition rule relative to a fixed frame. The increment $dR$ is constructively described by the matrices with obvious meanings. The process naturally incorporates angular velocity into the equation.

REFERENCES

[1] M. W. Spong, S. Hutchinson, and M. Vidyasagar, Robot Modeling and Control, John Wiley and Sons, 2006.

[2] L. Sciavicco and B. Siciliano, Modelling and Control of Robot Mnipulators, Springer, 2000.

[3] V. T. Coppola and N. H. McClamroch, "Spacecraft Attitude Control," in The Control Handbook, edited by W. Levine, CRC Press and IEEE Press, 1996, pp. 1303-1315.

[4] Rotation group SO(3), http://en.wikipedia.org/w/index.php?title=Rotation_group_SO(3)&oldid=553685388 , accessed on July 13, 2013.

[5] H. Asada and J.-J. E. Slotine, Robot Analysis and Control, John Wiley and Sons, 1986.

BIOGRAPHY

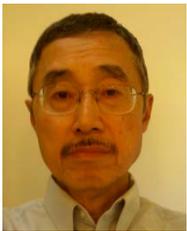

**Fumio Hamano** received the B.E. degree and the M.S.E. degree both in control engineering from Tokyo Institute of Technology, Tokyo, in 1973 and in 1975, respectively, and the Ph.D. degree in electrical engineering from University of Florida in 1979. Since 1989, he has been with the Electrical Engineering Department, California State University, Long Beach, where he is currently a Professor. He was chair of the department from 1996 to 2005. He was an associate professor in the ECE Department at the Florida Atlantic University through 1989, a visiting professor at the Department of Electrical Engineering, Computer Science, and Systems, University of Bologna, Italy, in 1994, and a lecturer for the NATO Advanced Study Institute on Expert Systems and Robotics (1990). His areas of interest are control theory and its applications, robotics, autonomous systems, and computer vision. He was a PI in various funded projects such as robot control, robot guidance with vision, robot end-effector integration, and machine vision algorithm development for industrial inspection and solar sub-system analysis and development.